\documentclass{paper}
\usepackage[utf8]{inputenc}
\usepackage{graphicx}
\usepackage{lineno}
\usepackage{subcaption}
\usepackage{natbib}

\title{UK COVID-19 Lockdown: What are the impacts on air pollution}
\author{J.E. Higham, M.A. Green \& C. Acosta Ram\'irez }

\begin{document}

\maketitle

\section{Abstract}

A country-wide `lock-down' imposed on the 23$^{rd}$ March 2020 in the UK had a significant impact on the UK's anthropogenic movements. The closure of work-places and restrictions imposed on visiting friends and family has radically reduced the amount of traffic on the roads. In this short communication, we use data from UK air-quality sensors to quantify air pollution trends pre- and post-lock-down. While we detect large falls in nitrogen dioxide at levels not seen over the last decade, trends in other pollutants were mixed especially when compared to historical data. It suggests that the implication that lock-down was beneficial for the environment was not so obvious.

\section{Introduction}
\noindent The global pandemic emerging from a novel strain of severe acute respiratory syndrome coronavirus 2 (SARS-CoV-2; Covid-19) has ravaged the world in 2020. This airborne infectious disease has shown to have a significant effect on mortality especially in vulnerable groups such as the elderly, those with cardiovascular disease, diabetes, respiratory diseases and cancer \citep{huang2020clinical}. Covid-19 has placed considerable social, economic and economic burdens on populations globally and will likely cause an economic crash similar to that seen in 1929 \citep{huang2020clinical,world2020novel,baker2020unprecedented}. While national governments have responded in different ways, most have introduced some form of social distancing measures to slow the spread of the infectious disease. On the 23$^{rd}$ March 2020, the United Kingdom (UK) government imposed a country-wide `lock-down' which introduced some of the most stringent social restrictions witnessed since World War II. Non-essential businesses, shops and services were asked to close. Individuals were asked to remain indoors at all times other than to shop for basic necessities, undertake daily exercise, access medical services, and travelling to and from work, but only when this couldn't be done from home. 

The extensive upheaval of society is likely to have significant changes to our environments particularly air quality. It may be hypothesised that with fewer individuals travelling outside and businesses closing, vehicle emissions will have fallen leading to improved air quality. The UK Department for Transport (DfT) has reported a 39\% decrease in all vehicle travel between 23 March and 28 April. However, this number had already been decreasing prior to the lock-down reducing from normal levels by 30\% on March 16 up until `lock-down' \citep{dft}. Recent studies \citep{muhammad2020covid,dantas2020impact,tobias2020changes,mahato2020effect} of China, France, Spain, India the USA and Brazil have shown that the effect of lock-down on their countries has had significant impacts in reducing the production of primarily Nitrogen Dioxide across all of these countries. 

Air-pollution is a substantial global issue, which according to the World Health Organization (WHO) is responsible for 3.7 million deaths annually \citep{who2014}. According to Public Health England (PHE) air-pollution is the biggest environmental threat to health in the UK, with between 28,000 and 36,000 deaths a year attributed to long-term exposure. The majority of this pollutants are not visible to the human eye, whilst it is possible to obtain large snapshot view of global pollution concentrations using satellites, these measurements have a very low temporal resolution (monthly) and only describe atmospheric concentrations \citep{martin2008satellite}. The most accurate way to determine ground-level air quality is by using single-point sensors \citep{guo2017estimating}. The UK's Department for Environment, Food and Rural Affairs (DEFRA) Automatic Urban and Rural Network (AURN) monitors has a large network of ~300 sensors distributed across the UK. The locations of these sensors are clustered around densely populated urban areas, with a higher spatial coverage around major cities. All measurements meet the required European Standards as set out in the European Ambient Air Quality Directive (2008/50/EC) measuring a combination of Nitrogen Dioxide, Sulphur Dioxide, PM2.5 and PM10 at hourly / sub-hourly intervals. 

Nitrogen dioxide (NO$_2$) is a highly reactive gas which pollutes our air mainly as a result of road traffic and energy production. In the UK in 2018, approximately 3 million tonnes of nitrogen dioxide was produced: 31\% of nitrogen dioxide was produced from road traffic; 20\% from energy production; 18\% from manufacturing; 14\% from other modes of transport; and, 17\% from other sources. Nitrogen dioxide is a considered a highly dangerous gas and has been associated with mild respiratory symptoms even at low concentrations \citep{chen2007outdoor}, while few studies have failed to reach a general consensus on the full effects nitrogen dioxide can have on the body the majority show that any exposure can affect mortality rates and lead to cardiovascular problems \citep{esplugues2011outdoor,maheswaran2012outdoor,hesterberg2009critical} and can even adverse effect on young children's development \citep{pershagen1995air}. 

Sulphur Dioxide (SO$_2$) is formed when fuel containing sulphur such as coil and oil typically created from energy production and manufacturing. In the UK sulphur dioxide production has been steadily decreasing with the UK's decreased reliance on coal-powered power stations. In 2018 the UK produced approximately 0.16 million tones of sulphur dioxide with 30\% being produced by energy industries 25\% domestic heating, 24\% manufacturing and 21\% from other sources. Exposure to sulphur dioxide has been associated with a number of symptoms including dyspnea (shortness of breath) and cough \citep{pikhart2001outdoor}. Epidemiological studies have reported exposure to elevated sulphur dioxide can contribute to an increase in mortality due to these respiratory and cardiovascular causes \citep{tertre2002short}; even short term exposure to higher concentrations leading to hospitalisation of the elderly and vulnerable \citep{martins2002air}.

Particulate matter classified by the size of 2.5 and 10 (PM2.5 and PM10) microns are two of the most commonly used in air quality assessments. Particle matter is defined as everything in the air that isn't a gas, they originate from a number different sources including domestic heating, industry and vehicles and are made up from a variety of different chemical compounds. In the UK in 2018 domestic combustion was a major source of PM emissions, accounting for 27\% and 44\% of PM10 and PM2.5 respectively, industrial processing and accounted for 43\% of PM10 and 29\% of PM2.5, vehicles 11\% of both PM10 and PM2.5 and the rest from other sources. In 2018 the UK produced 0.12 million tonnes of PM2.5 and 0.18 of PM10. Particle matter as shown to cause respiratory problems \citep{pope2006health} and smaller PM2.5 have been proved to be easily absorbed into the bloodstream, causing hardening of arteries leading to strokes and cardiovascular problems \citep{santibanez2013five} with studies also linking PM2.5 and SO2 causing lung cancer \citep{xing2019spatial} PM2.5 and SO2.

Clearly, all forms of air pollutants can have serious implications to human health. Importantly, their risk to respiratory health, as well as interactions with chronic co-morbidities, means air pollution can exacerbate outcomes related to Covid-19 and may place individuals at higher risk of mortality. The aim of our study is to explore how trends in air pollutants have changed in the UK before and after the introduction of the lock-down. we hypothesise that emissions of all major air pollutants have fallen since 23$^{rd}$ March.

\section{Results \& Discussion}

\begin{figure}
    \centering
    \includegraphics[width=1.1\textwidth]{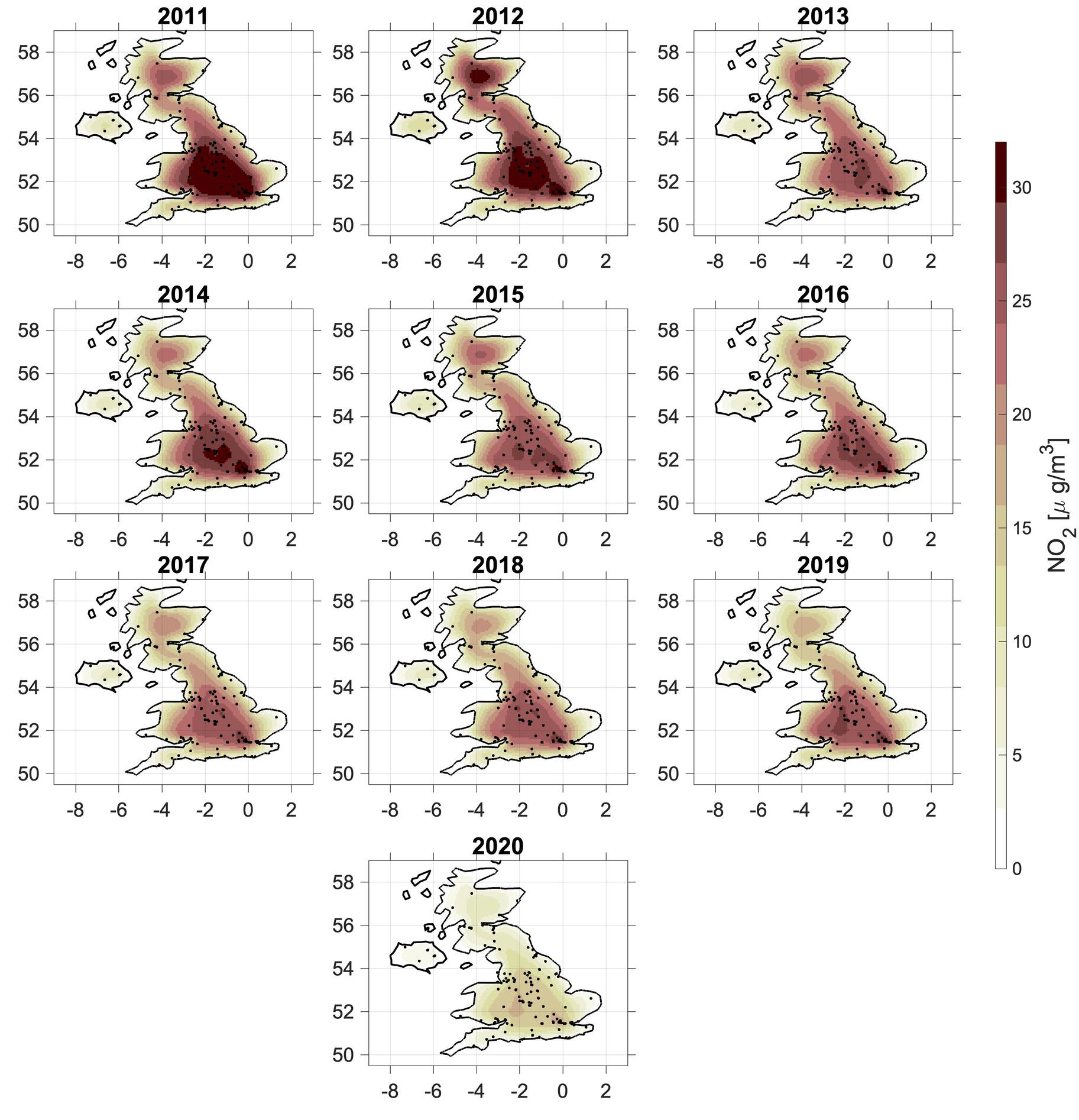}
    \caption{Figures showing time-average contours of NO$_2$ pollution in period of `lockdown" (23$^{rd}$ March 2020 - 28$^{th}$ April 2020) compared with same period in previous 9 years}
    \label{fig:no2contour}
\end{figure}

\begin{figure}
    \centering
    \includegraphics[width=1.1\textwidth]{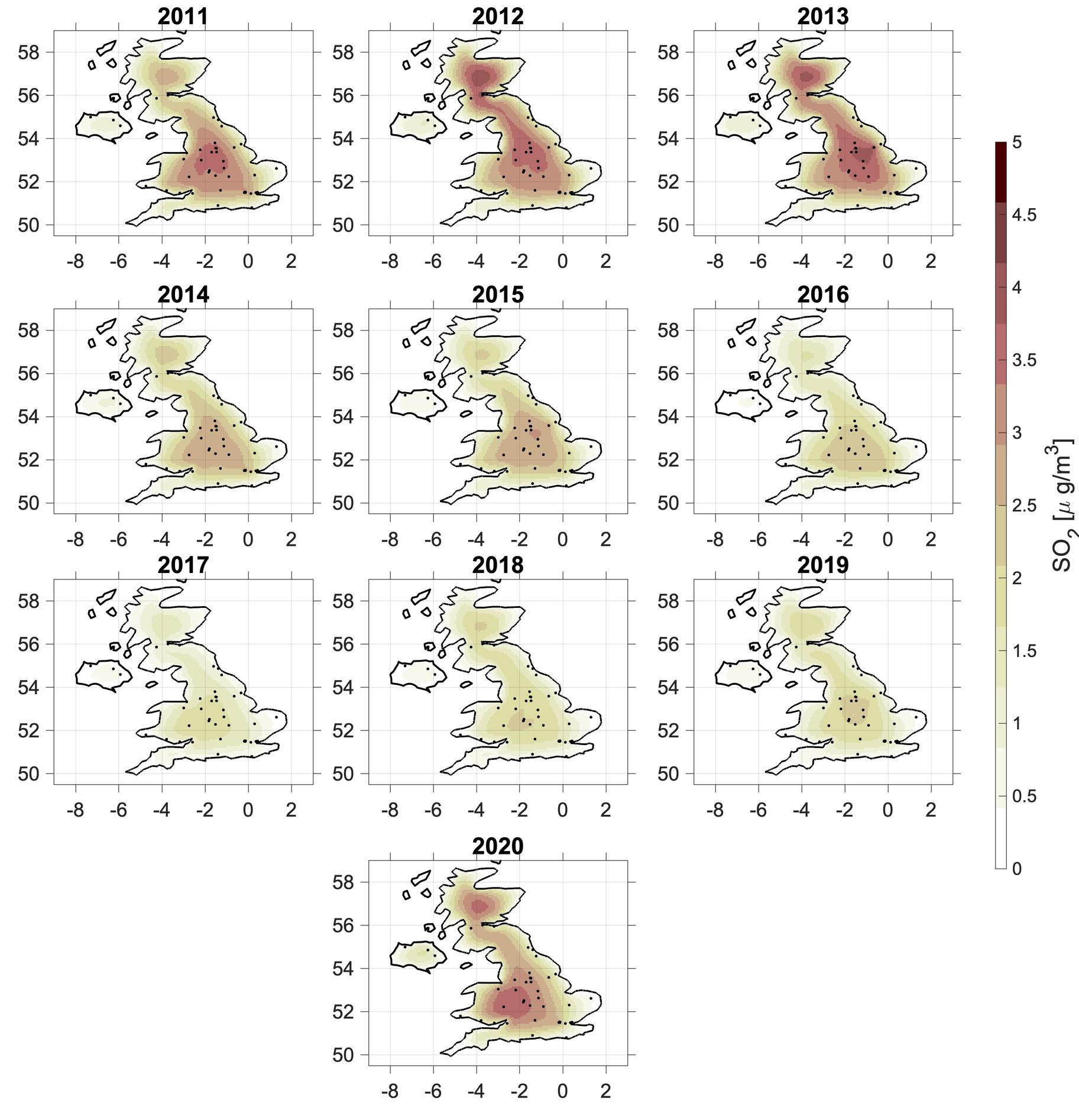}
    \caption{Figures showing time-average contours of SO$_2$ pollution in period of `lock-down' (23$^{rd}$ March 2020 - 28$^{th}$ April 2020) compared with same period in previous 9 years}
    \label{fig:so2contour}
\end{figure}

\begin{figure}
    \centering
    \includegraphics[width=1.1\textwidth]{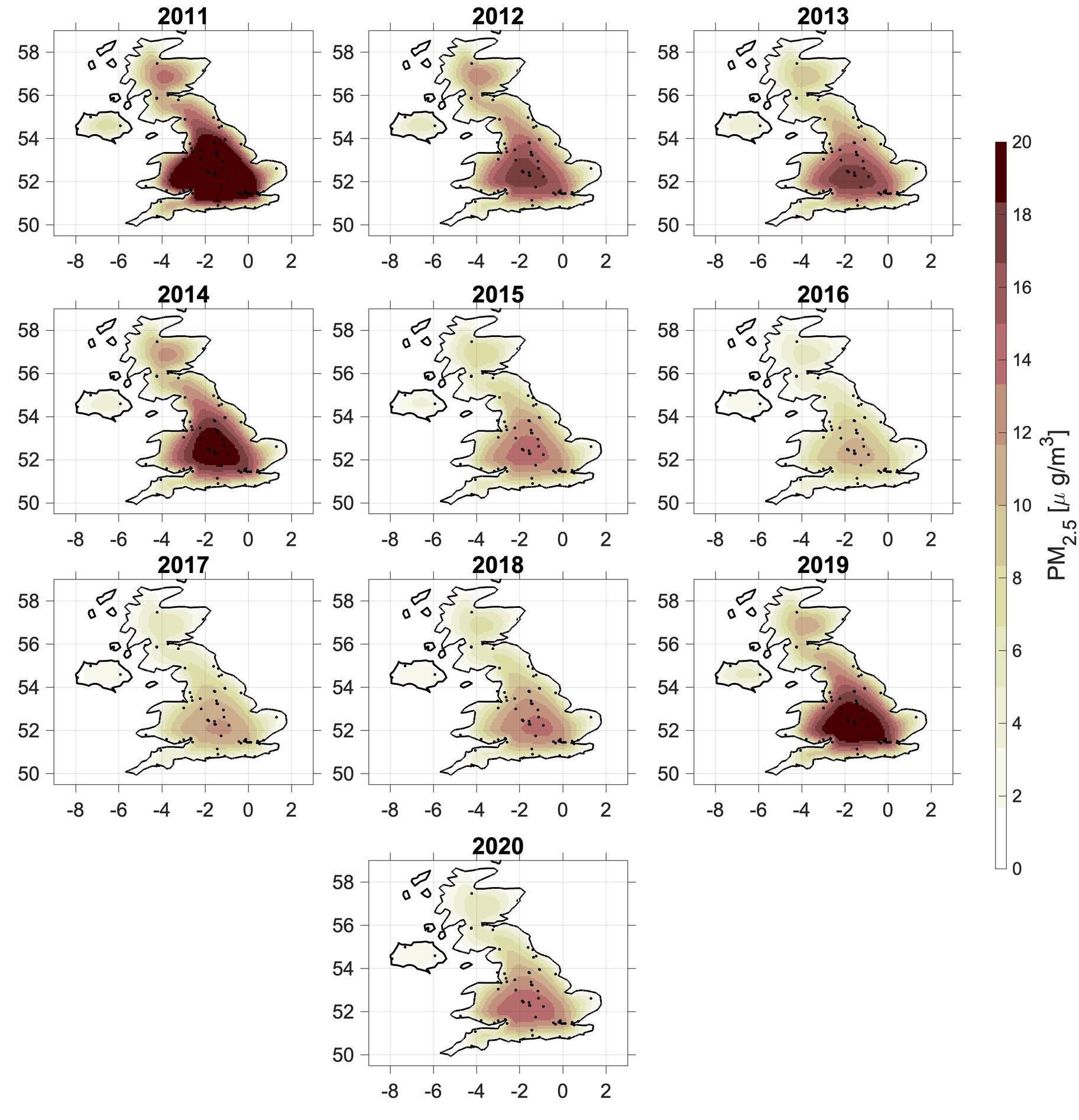}
    \caption{Figures showing time-average contours of PM2.5 pollution in period of `lock-down' (23$^{rd}$ March 2020 - 28$^{th}$ April 2020) compared with same period in previous 9 years}
    \label{fig:pm25contour}
\end{figure}

\begin{figure}
    \centering
    \includegraphics[width=1.1\textwidth]{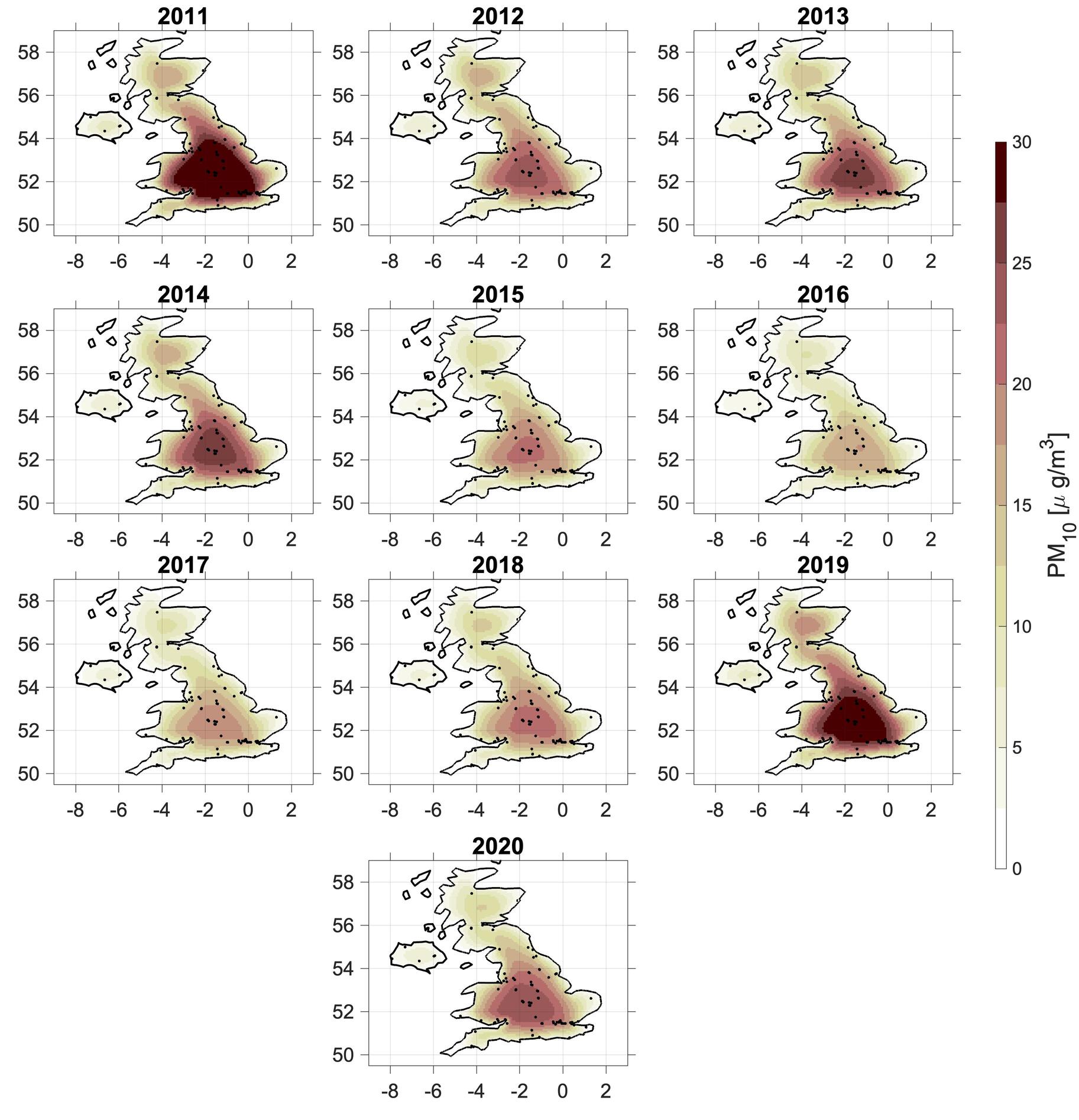}
    \caption{Figures showing time-average contours of PM10 pollution in period of `lock-down' (23$^{rd}$ March 2020 - 28$^{th}$ April 2020) compared with same period in previous 9 years}
    \label{fig:pm10contour}
\end{figure}

\begin{figure}
    \centering
    \includegraphics[width=\textwidth]{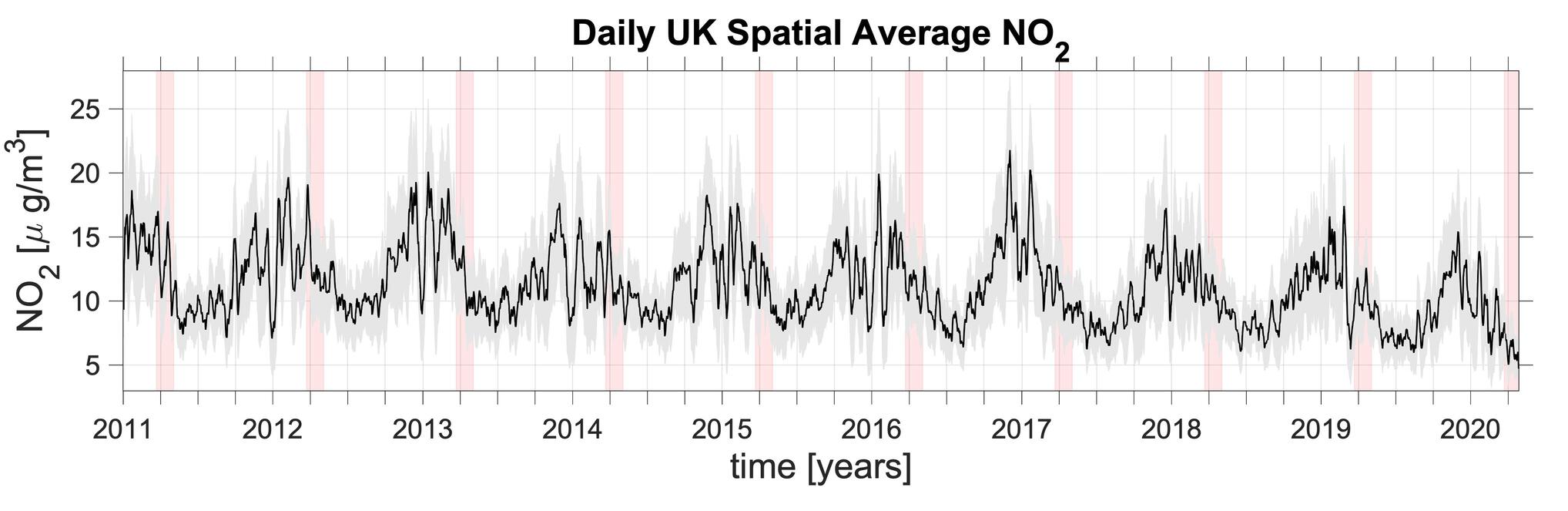}
    \includegraphics[width=\textwidth]{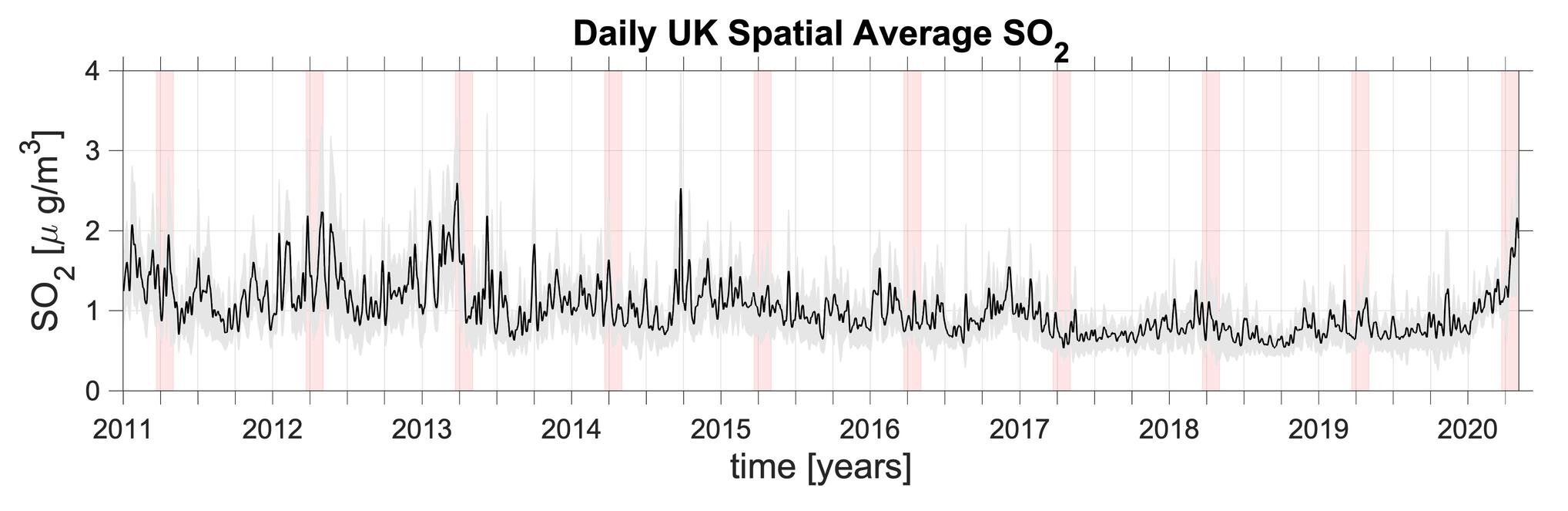}
    \includegraphics[width=\textwidth]{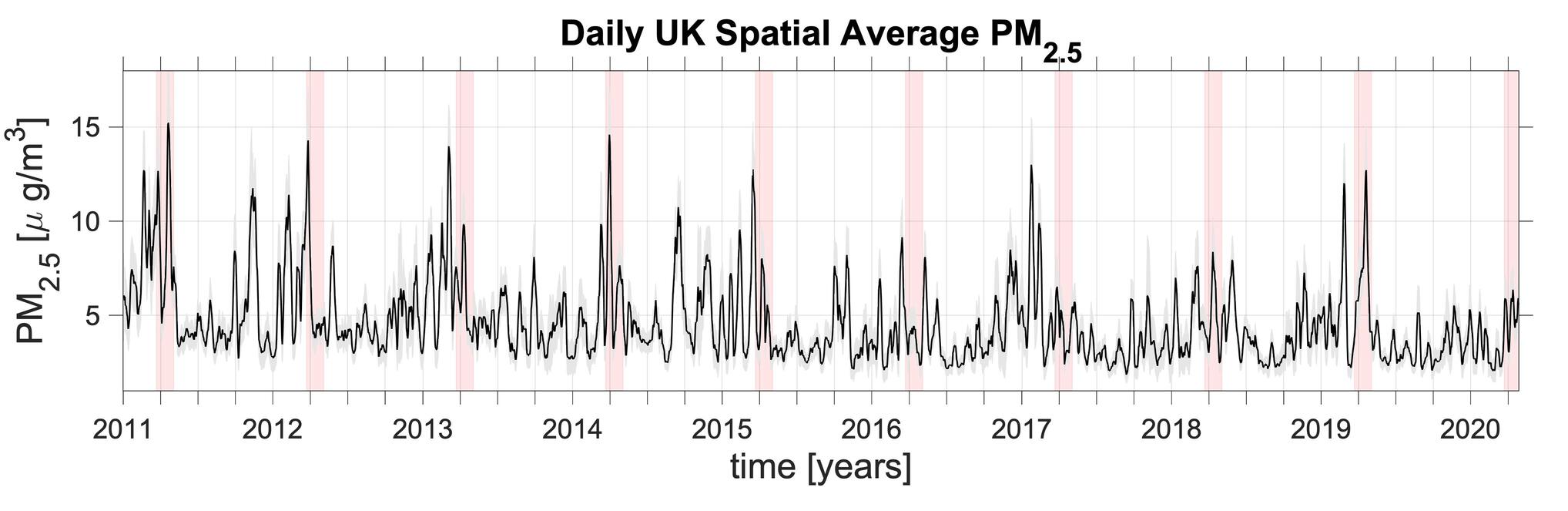}
    \includegraphics[width=\textwidth]{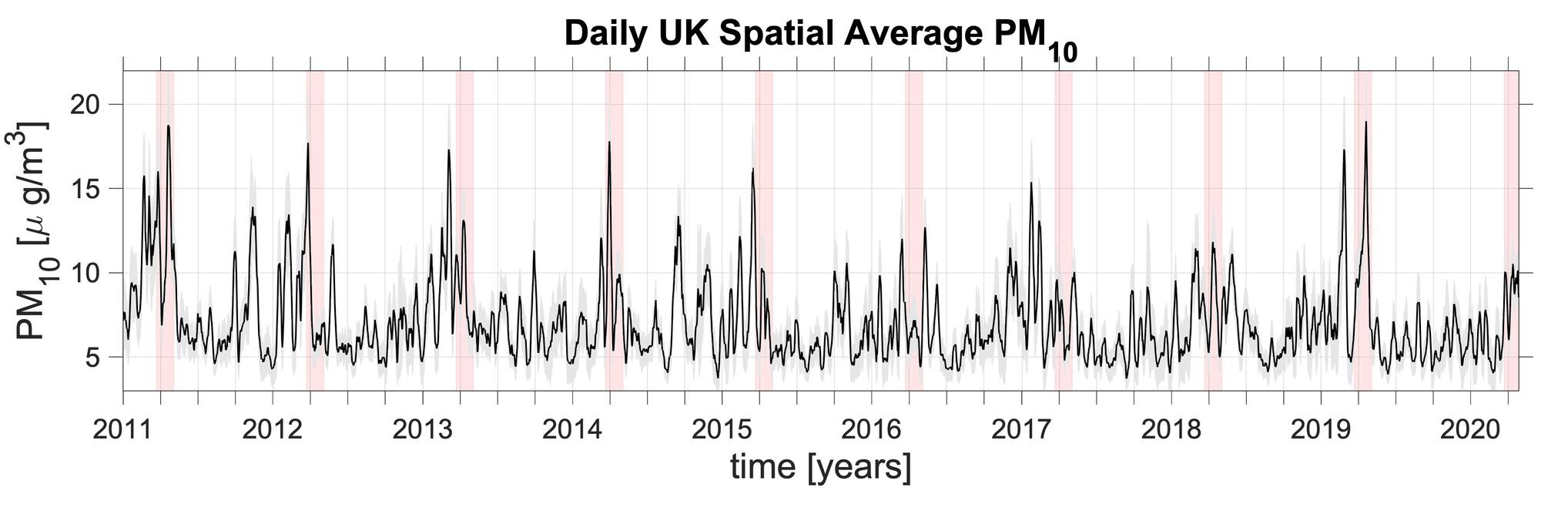}
    \caption{Spatial daily average time-series of past ten years UK air-pollution, with daily variance highlighted in grey. First row Nitrogen Dioxide, second row Sulphur Dioxide, third row PM2.5 bottom row PM10. Red highlighted areas denote period of `lock-down' (previous years highlighted for comparison)}
    \label{fig:yearly}
\end{figure}

\begin{figure}
    \centering
    \includegraphics[width=\textwidth]{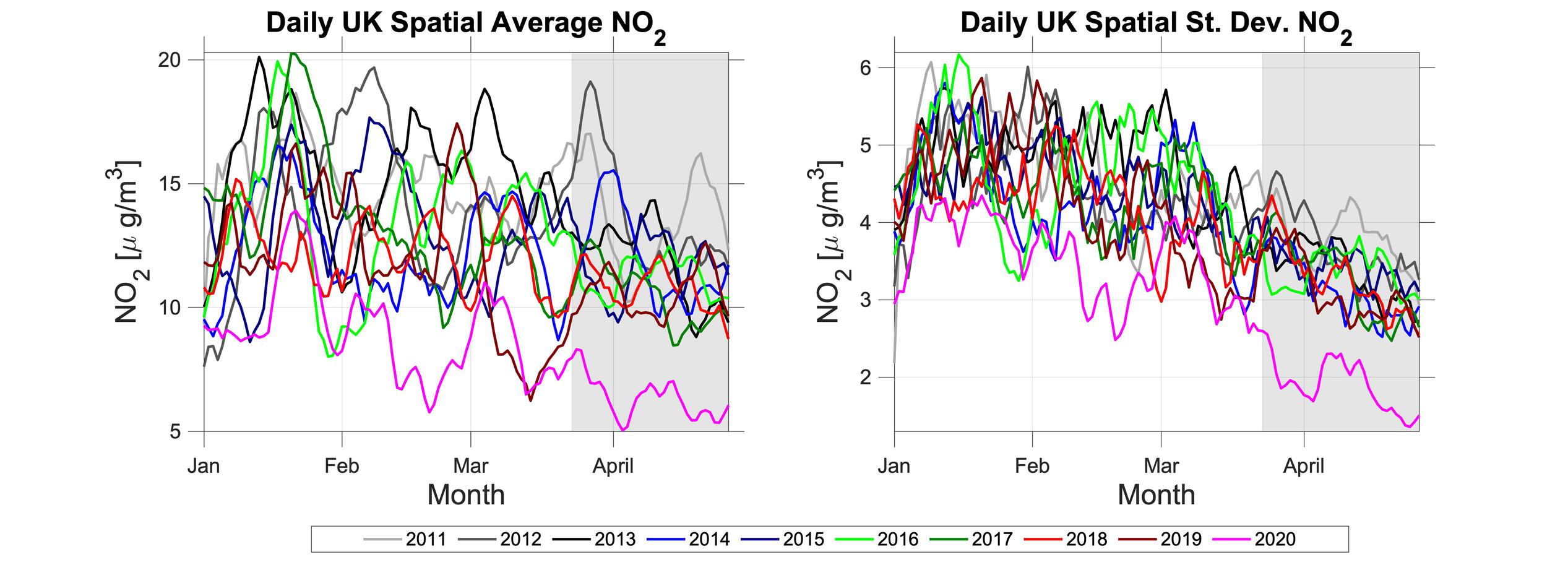} 
    \includegraphics[width=\textwidth]{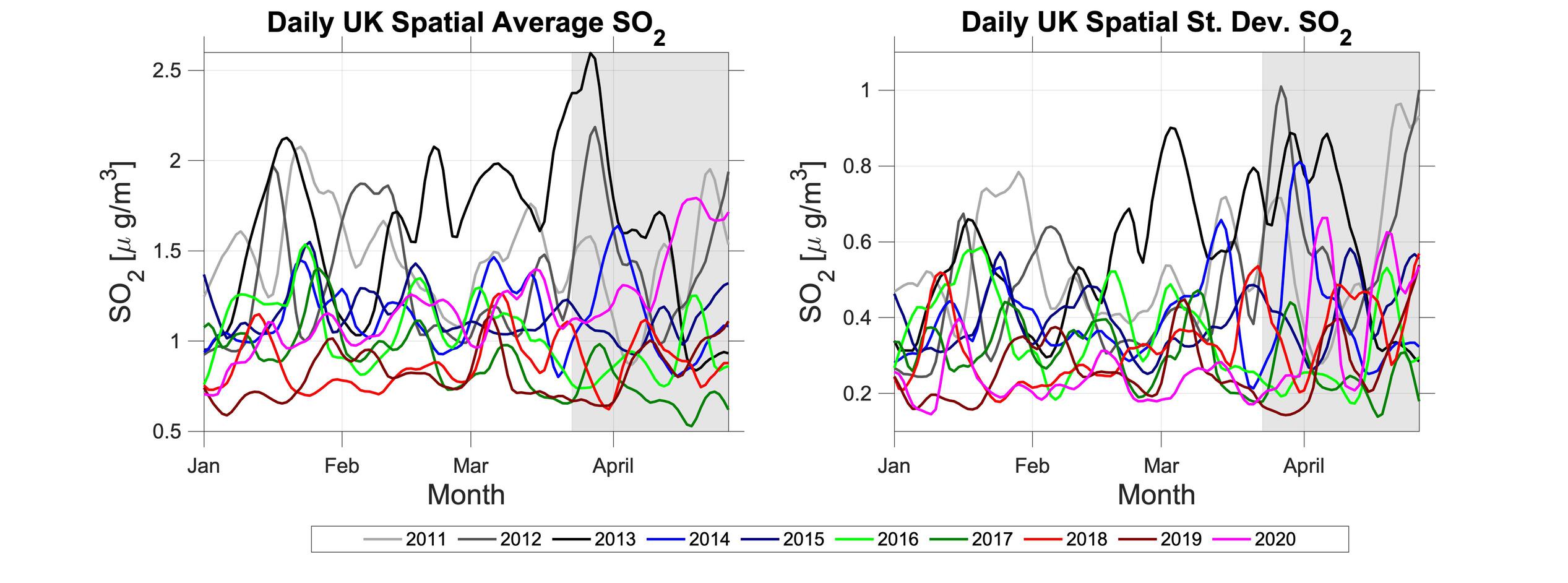} 
    \includegraphics[width=\textwidth]{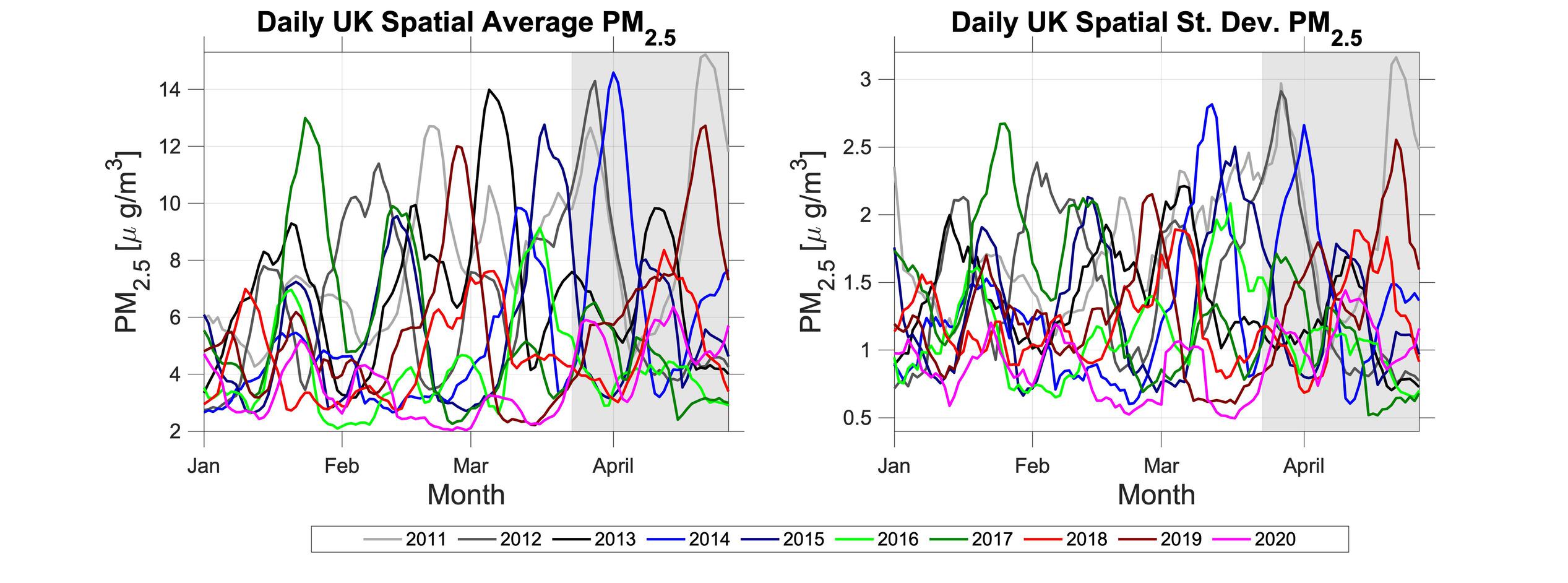}
    \includegraphics[width=\textwidth]{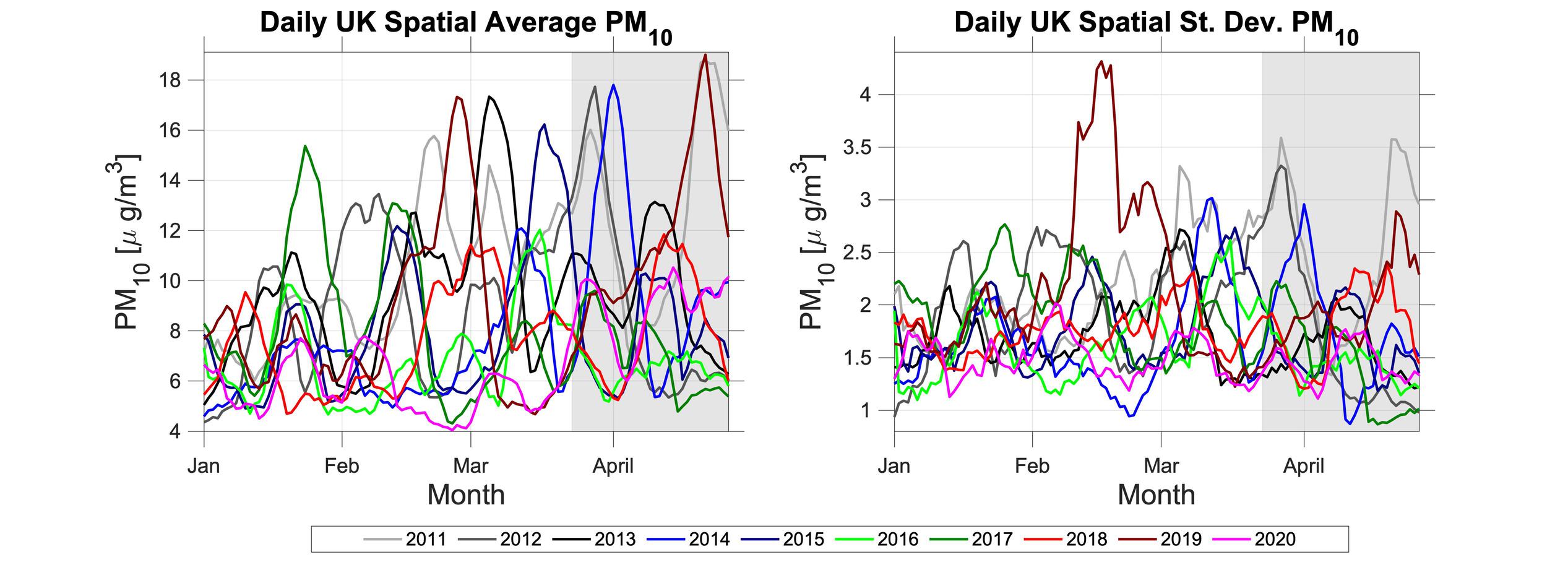}
    \caption{Spatial daily average time-series (left) and spatial daily standard deviation (right) of first four months of the last ten years of UK's air-pollution. First row Nitrogen Dioxide, second row Sulphur Dioxide, third row PM2.5 bottom row PM10. Grey highlighted area denotes `lock-down' period.}
    \label{fig:monthly}
\end{figure}

\noindent Figures~\ref{fig:no2contour}-\ref{fig:pm10contour} show a series of contour plots created from the points the DEFRA AURN stations. The point time-series data is gridded using a multi-resolution, cubic, gridded-interpolator \citep{higham2019and}. The method assumes an adaptive grid to distribute measurements ensuring at least one point is described by one grid cell \citep{liszka1984interpolation}. In each of the figures the black dots signify the locations of the air-quality sensors (please note that for different air-pollutants there as a differing number of data points). In each of the figures the mean value is determined between the date of 23$^{rd}$ March and 28$^{th}$ April for the past 10 years. A sample of 10 years is used to ensure there is a significant change caused by the `lock-down'. In figs.~\ref{fig:yearly} we present the daily average measurement for the past ten years, we also present the mean plus or minus the daily standard deviations (filled light grey area), this variance represents the average daily production of the pollutants. The red highlighted areas show the regions used to create the ensemble-averaged presented in Figs.~\ref{fig:no2contour}-\ref{fig:pm25contour}. Figures.~\ref{fig:monthly} focus in on 2020, comparing UK spatial average time-series with the previous 9 years. Here the `lock-down' phase is highlighted in grey, in the left-hand panes we present the daily averages and on the right-hand pane the daily standard deviations.   

As expected from the contour plots and yearly time average we can see over the past 10 years there has been a general decrease in the production of all air pollutants. This is likely to be attributed to the declining use in fossil fuels. We also see that near heavily densely populated cities, such as London and Birmingham, on average we see higher Nitrogen dioxide, Sulphur dioxide levels, and generally in the UK England is the most polluted. We also see high PM2.5 levels in Scotland, as the averages taken are during periods of colder weather this is likely attributed to domestic heating in the cooler northern regions of the UK. From the contour plots we also see results echoing those in the studies of \citep{muhammad2020covid,dantas2020impact,tobias2020changes,mahato2020effect} we see, during the `lock-down' period, there is a significant average decrease in the production of Nitrogen dioxide. This contrasts with levels around London and Birmingham, whilst lower in comparison with other years, remain slightly elevated during the lock-down. The spatial average of these results presented in Fig.~\ref{fig:yearly} highlight compared with previous years there is a significant drop. By focusing in on the period leading up to the `lock-down' in Fig.~\ref{fig:monthly} these changes are quite evident with levels reducing on average by approximately one third and daily production almost halving. These findings reflect the 69\% decrease in road traffic as reported by the DfT. 

Notably from all of the figures we see there is no real significant impact of `lock-down' on the production of PM2.5 nor PM10. Figures~\ref{fig:monthly} might suggest there is a slight decrease in the production of PM2.5 and PM10, however as discussed by \citep{zhang2016exploring} the complex make up of these pollutants make such signals difficult to interpret and a number of factors such as increased humidity \citep{cheng2015humidity} and dust storms \citep{vieno2016uk,tao2013chemical} can easily distort these readings. Similarly, the on-average 2.5 degrees Celsius increase in temperature during this period in 2020, compared with 2019, might account for a reduction in domestic heating. 

Most interestingly, and perhaps unexpectedly, preceding and during the `lock-down' period we see that there is an increase in the production of Sulphur Dioxide; Fig~\ref{fig:yearly} even shows there is a five year high. Fig~\ref{fig:so2contour} shows an increase in concentration surrounding Birmingham, London and central Scotland. Figure~\ref{fig:monthly} shows, although there is an average increase the daily standard deviation is comparable with previous years. This perhaps suggests there is an increase in the background of Sulphur dioxide, apposed to daily cycles in production. This could be attributed to several sources including wind speed, humidity and climatic changes \citep{van1990effect}, although these could also be attributed to increased human cremation as suggested by an older study \citep{qian2007short} conducted in Wuhan, China.

\section{Conclusions}

\noindent To our knowledge, our study presents the first analysis of air quality trends in the UK. We demonstrate a mixed and complex picture. There has been a substantial fall in nitrogen dioxide levels observed consistently across the UK and at an unprecedented level. However, similar trends were not observed for sulphur dioxide which has increased over the same period. Trends for PM2.5 and PM10 suggest that recent declines are not significantly different from historical trends. 

Improved environments through less nitrogen dioxide may have public health benefits both in short and longer-term - similarly, climate change emissions too. Renewed interest in environments and seeing cleaner air might result in greater future action unless people use `non-lock-down' for travelling more to account for their current comprised situations. An increase in sulphur dioxide is a worrying finding, whilst at present current levels remain safe, further rises may require action. 

In this short communication that we do not look in detail at the drivers of trends, rather just present the key findings. This will be the feature of a future in-depth study where we will connect the trends in air quality to population groups (e.g. inequalities by the level of deprivation), including how they may affect responses to Covid-19 (air pollution appears to be a factor in experiences and deaths - affecting respiratory health). 

\section{Acknowledgements}
\noindent No external/internal funding has been received for this work. 

\bibliography{bib.bib}

\begin{thebibliography}{30}
\providecommand{\natexlab}[1]{#1}
\providecommand{\url}[1]{\texttt{#1}}
\expandafter\ifx\csname urlstyle\endcsname\relax
  \providecommand{\doi}[1]{doi: #1}\else
  \providecommand{\doi}{doi: \begingroup \urlstyle{rm}\Url}\fi

\bibitem[who(2014)]{who2014}
{WHO}: {S}even million premature deaths annually linked to air pollution, March
  2014.
\newblock URL
  \url{https://www.who.int/mediacentre/news/releases/2014/air-pollution/en/}.

\bibitem[dft(2020)]{dft}
{UK} {D}epartment for {T}ransport {S}tatistics, April 2020.
\newblock URL
  \url{https://www.gov.uk/government/organisations/department-for-transport/about/statistics}.

\bibitem[Baker et~al.(2020)Baker, Bloom, Davis, Kost, Sammon, and
  Viratyosin]{baker2020unprecedented}
Scott~R Baker, Nicholas Bloom, Steven~J Davis, Kyle Kost, Marco Sammon, and
  Tasaneeya Viratyosin.
\newblock The unprecedented stock market reaction to {COVID-19}.
\newblock \emph{Covid Economics: {V}etted and Real-Time Papers}, 1\penalty0
  (3), 2020.

\bibitem[Chen et~al.(2007)Chen, Kuschner, Gokhale, and Shofer]{chen2007outdoor}
Tze-Ming Chen, Ware~G Kuschner, Janaki Gokhale, and Scott Shofer.
\newblock Outdoor air pollution: nitrogen dioxide, sulfur dioxide, and carbon
  monoxide health effects.
\newblock \emph{The American journal of the medical sciences}, 333\penalty0
  (4):\penalty0 249--256, 2007.

\bibitem[Cheng et~al.(2015)Cheng, He, Du, Zheng, Duan, and
  Ma]{cheng2015humidity}
Yuan Cheng, Ke-bin He, Zhen-yu Du, Mei Zheng, Feng-kui Duan, and Yong-liang Ma.
\newblock Humidity plays an important role in the {PM}2. 5 pollution in
  {B}eijing.
\newblock \emph{Environmental pollution}, 197:\penalty0 68--75, 2015.

\bibitem[Dantas et~al.(2020)Dantas, Siciliano, Fran{\c{c}}a, da~Silva, and
  Arbilla]{dantas2020impact}
Guilherme Dantas, Bruno Siciliano, Bruno~Boscaro Fran{\c{c}}a, Cleyton~M
  da~Silva, and Graciela Arbilla.
\newblock The impact of covid-19 partial lockdown on the air quality of the
  city of {R}io de {J}aneiro, {B}razil.
\newblock \emph{Science of The Total Environment}, 729:\penalty0 139085, 2020.

\bibitem[Esplugues et~al.(2011)Esplugues, Ballester, Estarlich, Llop,
  Fuentes-Leonarte, Mantilla, Vioque, and I{\~n}iguez]{esplugues2011outdoor}
Ana Esplugues, Ferran Ballester, Marisa Estarlich, Sabrina Llop, Virginia
  Fuentes-Leonarte, Enrique Mantilla, Jes{\'u}s Vioque, and Carmen I{\~n}iguez.
\newblock Outdoor, but not indoor, nitrogen dioxide exposure is associated with
  persistent cough during the first year of life.
\newblock \emph{Science of the total environment}, 409\penalty0 (22):\penalty0
  4667--4673, 2011.

\bibitem[Guo et~al.(2017)Guo, Tang, Gong, and Zhang]{guo2017estimating}
Yuanxi Guo, Qiuhong Tang, Dao-Yi Gong, and Ziyin Zhang.
\newblock Estimating ground-level {PM}2. 5 concentrations in {B}eijing using a
  satellite-based geographically and temporally weighted regression model.
\newblock \emph{Remote Sensing of Environment}, 198:\penalty0 140--149, 2017.

\bibitem[Hesterberg et~al.(2009)Hesterberg, Bunn, McClellan, Hamade, Long, and
  Valberg]{hesterberg2009critical}
Thomas~W Hesterberg, William~B Bunn, Roger~O McClellan, Ali~K Hamade,
  Christopher~M Long, and Peter~A Valberg.
\newblock Critical review of the human data on short-term nitrogen dioxide
  ({NO2}) exposures: {E}vidence for {NO2} no-effect levels.
\newblock \emph{Critical reviews in toxicology}, 39\penalty0 (9):\penalty0
  743--781, 2009.

\bibitem[Higham and Brevis(2019)]{higham2019and}
JE~Higham and W~Brevis.
\newblock When, what and how image transformation techniques should be used to
  reduce error in {P}article {I}mage {V}elocimetry data?
\newblock \emph{Flow Measurement and Instrumentation}, 66:\penalty0 79--85,
  2019.

\bibitem[Huang et~al.(2020)Huang, Wang, Li, Ren, Zhao, Hu, Zhang, Fan, Xu, Gu,
  et~al.]{huang2020clinical}
Chaolin Huang, Yeming Wang, Xingwang Li, Lili Ren, Jianping Zhao, Yi~Hu,
  Li~Zhang, Guohui Fan, Jiuyang Xu, Xiaoying Gu, et~al.
\newblock Clinical features of patients infected with 2019 novel coronavirus in
  {W}uhan, {C}hina.
\newblock \emph{The lancet}, 395\penalty0 (10223):\penalty0 497--506, 2020.

\bibitem[Liszka(1984)]{liszka1984interpolation}
Tadeusz Liszka.
\newblock An interpolation method for an irregular net of nodes.
\newblock \emph{International Journal for Numerical Methods in Engineering},
  20\penalty0 (9):\penalty0 1599--1612, 1984.

\bibitem[Mahato et~al.(2020)Mahato, Pal, and Ghosh]{mahato2020effect}
Susanta Mahato, Swades Pal, and Krishna~Gopal Ghosh.
\newblock Effect of lockdown amid {COVID-19} pandemic on air quality of the
  megacity {D}elhi, {I}ndia.
\newblock \emph{Science of The Total Environment}, page 139086, 2020.

\bibitem[Maheswaran et~al.(2012)Maheswaran, Pearson, Smeeton, Beevers,
  Campbell, and Wolfe]{maheswaran2012outdoor}
Ravi Maheswaran, Tim Pearson, Nigel~C Smeeton, Sean~D Beevers, Michael~J
  Campbell, and Charles~D Wolfe.
\newblock Outdoor air pollution and incidence of ischemic and hemorrhagic
  stroke: {A} small-area level ecological study.
\newblock \emph{Stroke}, 43\penalty0 (1):\penalty0 22--27, 2012.

\bibitem[Martin(2008)]{martin2008satellite}
Randall~V Martin.
\newblock Satellite remote sensing of surface air quality.
\newblock \emph{Atmospheric environment}, 42\penalty0 (34):\penalty0
  7823--7843, 2008.

\bibitem[Martins et~al.(2002)Martins, de~Oliveira~Latorre, Saldiva, and
  Braga]{martins2002air}
Lourdes~Concei{\c{c}}{\~a}o Martins, Maria Ros{\'a}rio~Dias
  de~Oliveira~Latorre, Paulo Hil{\'a}rio~Nascimento Saldiva, and Alf{\'e}sio
  Lu{\'\i}s~Ferreira Braga.
\newblock Air pollution and emergency room visits due to chronic lower
  respiratory diseases in the elderly: {A}n ecological time-series study in
  {S}ao {P}aulo, {B}razil.
\newblock \emph{Journal of occupational and environmental medicine},
  44\penalty0 (7):\penalty0 622--627, 2002.

\bibitem[Muhammad et~al.(2020)Muhammad, Long, and Salman]{muhammad2020covid}
Sulaman Muhammad, Xingle Long, and Muhammad Salman.
\newblock {COVID-19} pandemic and environmental pollution: {A} blessing in
  disguise?
\newblock \emph{Science of The Total Environment}, page 138820, 2020.

\bibitem[PERSHAGEN et~al.(1995)PERSHAGEN, Rylander, Norberg, Eriksson, and
  Nordvall]{pershagen1995air}
G{\"O}RAN PERSHAGEN, Emma Rylander, Staffan Norberg, Margareta Eriksson, and
  S~Lennart Nordvall.
\newblock Air pollution involving nitrogen dioxide exposure and wheezing
  bronchitis in children.
\newblock \emph{International journal of epidemiology}, 24\penalty0
  (6):\penalty0 1147--1153, 1995.

\bibitem[Pikhart et~al.(2001)Pikhart, Bobak, Gorynski, Wojtyniak, Danova,
  Celko, Kriz, Briggs, and Elliott]{pikhart2001outdoor}
Hynek Pikhart, Martin Bobak, Pawel Gorynski, Bogdan Wojtyniak, Jana Danova,
  Martin~A Celko, Bohumir Kriz, David Briggs, and Paul Elliott.
\newblock Outdoor sulphur dioxide and respiratory symptoms in {C}zech and
  {P}olish school children: a small-area study (saviah).
\newblock \emph{International archives of occupational and environmental
  health}, 74\penalty0 (8):\penalty0 574--578, 2001.

\bibitem[Pope~III and Dockery(2006)]{pope2006health}
C~Arden Pope~III and Douglas~W Dockery.
\newblock Health effects of fine particulate air pollution: {L}ines that
  connect.
\newblock \emph{Journal of the air \& waste management association},
  56\penalty0 (6):\penalty0 709--742, 2006.

\bibitem[Qian et~al.(2007)Qian, He, Lin, Kong, Liao, Yang, Bentley, and
  Xu]{qian2007short}
Zhengmin Qian, Qingci He, Hung-Mo Lin, Lingli Kong, Duanping Liao, Niannian
  Yang, Christy~M Bentley, and Shuangqing Xu.
\newblock Short-term effects of gaseous pollutants on cause-specific mortality
  in {W}uhan, {C}hina.
\newblock \emph{Journal of the Air \& Waste Management Association},
  57\penalty0 (7):\penalty0 785--793, 2007.

\bibitem[Santiba{\~n}ez et~al.(2013)Santiba{\~n}ez, Ibarra, Matus, Seguel,
  et~al.]{santibanez2013five}
Daniela~A Santiba{\~n}ez, Sergio Ibarra, Patricia Matus, Rodrigo Seguel, et~al.
\newblock A five-year study of particulate matter ({PM}2. 5) and
  cerebrovascular diseases.
\newblock \emph{Environmental Pollution}, 181:\penalty0 1--6, 2013.

\bibitem[Tao et~al.(2013)Tao, Zhang, Engling, Zhang, Yang, Cao, Zhu, Wang, and
  Luo]{tao2013chemical}
Jun Tao, Leiming Zhang, Guenter Engling, Renjian Zhang, Yihong Yang, Junji Cao,
  Chongshu Zhu, Qiyuan Wang, and Lei Luo.
\newblock Chemical composition of {PM}2. 5 in an urban environment in
  {C}hengdu, {C}hina: {I}mportance of springtime dust storms and biomass
  burning.
\newblock \emph{Atmospheric Research}, 122:\penalty0 270--283, 2013.

\bibitem[Tertre et~al.(2002)Tertre, Qu{\'e}nel, Eilstein, Medina, Prouvost,
  Pascal, Boumghar, Saviuc, Zeghnoun, Filleul, et~al.]{tertre2002short}
Alain~Le Tertre, Philippe Qu{\'e}nel, Daniel Eilstein, Sylvia Medina, Helene
  Prouvost, Laurence Pascal, Azzedine Boumghar, Philippe Saviuc, Abdelkrim
  Zeghnoun, Laurent Filleul, et~al.
\newblock Short-term effects of air pollution on mortality in nine {F}rench
  cities: {A} quantitative summary.
\newblock \emph{Archives of Environmental Health: An International Journal},
  57\penalty0 (4):\penalty0 311--319, 2002.

\bibitem[Tob{\'\i}as et~al.(2020)Tob{\'\i}as, Carnerero, Reche, Massagu{\'e},
  Via, Minguill{\'o}n, Alastuey, and Querol]{tobias2020changes}
Aurelio Tob{\'\i}as, Cristina Carnerero, Cristina Reche, Jordi Massagu{\'e},
  Marta Via, Mar{\'\i}a~Cruz Minguill{\'o}n, Andr{\'e}s Alastuey, and Xavier
  Querol.
\newblock Changes in air quality during the lockdown in {B}arcelona ({S}pain)
  one month into the {SARS-CoV-2} epidemic.
\newblock \emph{Science of The Total Environment}, page 138540, 2020.

\bibitem[Van~Hove et~al.(1990)Van~Hove, Vredenberg, and Adema]{van1990effect}
LWA Van~Hove, WJ~Vredenberg, and EH~Adema.
\newblock The effect of wind velocity, air temperature and humidity on {NH3}
  and {SO2} transfer into bean leaves (phaseolus vulgaris l.).
\newblock \emph{Atmospheric Environment. Part A. General Topics}, 24\penalty0
  (5):\penalty0 1263--1270, 1990.

\bibitem[Vieno et~al.(2016)Vieno, Heal, Twigg, MacKenzie, Braban, Lingard,
  Ritchie, Beck, M{\'o}ring, Ots, et~al.]{vieno2016uk}
Massimo Vieno, Mathew~R Heal, Marsailidh~M Twigg, IA~MacKenzie, Christine~F
  Braban, JJN Lingard, S~Ritchie, RC~Beck, A~M{\'o}ring, R~Ots, et~al.
\newblock The uk particulate matter air pollution episode of {M}arch--{A}pril
  2014: {M}ore than {S}aharan dust.
\newblock \emph{Environmental Research Letters}, 11\penalty0 (4):\penalty0
  044004, 2016.

\bibitem[(WHO) et~al.(2020)]{world2020novel}
World Health~Organization (WHO) et~al.
\newblock Novel coronavirus—{T}hailand (ex-china). 2020, 2020.

\bibitem[Xing et~al.(2019)Xing, Xu, Liao, Xing, Cheng, Hu, and
  Wang]{xing2019spatial}
DF~Xing, CD~Xu, XY~Liao, TY~Xing, SP~Cheng, MG~Hu, and JX~Wang.
\newblock Spatial association between outdoor air pollution and lung cancer
  incidence in {C}hina.
\newblock \emph{BMC public health}, 19\penalty0 (1):\penalty0 1377, 2019.

\bibitem[Zhang et~al.(2016)Zhang, Wang, and Zhang]{zhang2016exploring}
Haifeng Zhang, Zhaohai Wang, and Wenzhong Zhang.
\newblock Exploring spatiotemporal patterns of {PM}2. 5 in china based on
  ground-level observations for 190 cities.
\newblock \emph{Environmental Pollution}, 216:\penalty0 559--567, 2016.

\end{thebibliography}
\bibliographystyle{plainnat}

\end{document}